\documentclass{article}
\usepackage{amsmath}

\begin{document}
\vskip 1truein
\begin{center}
{\Large {\bf ON CONSISTENCE OF MATTER COUPLING IN \\ GL(3,R) GAUGE
FORMULATION OF GRAVITY }} \vskip 10pt {\bf ROLANDO
GAITAN}${}^{a, }${\footnote {e-mail:
rgaitan@fisica.ciens.ucv.ve}} \\
${}^a${\it Grupo de F\'{\i}sica Te\'orica, Departamento de F\'\i
sica, Facultad de Ciencias y Tecnolog\'\i a, \\
Universidad de
Carabobo,A.P. 129
Valencia 2001, Edo. Carabobo, Venezuela.}\\
\end{center}
\vskip .2truein
\begin{center}
\begin{minipage}{4.5in}\footnotesize\baselineskip=10pt
   \parindent=0pt
A covariant scheme for matter coupling with a $GL(3,R)$ gauge
formulation of gravity is studied. We revisit a known Yang-Mills
type construction, where quadratical power of cosmological
constant have to be considered in consistence with vacuum
Einstein's gravity. Then, matter coupling with gravity is
introduced and some constraints on fields and background appear.
Finally we elucidate that introduction of auxiliary fields
decreases the number of these constraints.
\end{minipage}
\end{center}
\vskip .2truein

Within a large list of gauge formulations for gravity, we focus
our attention in the well known idea of a construction based on
$GL(N,R)$ as a gauge group$^{1}$, because its simplicity. Here, is
expected that a Yang-Mills lagrangian density will
be related to a quadratic lagrangian density on Riemann-Christoffel 
curvature.
This type of lagrangians have a great interest, because these
yield theories where the renormalization problems are much less
severe$^{2}$, among other things.

The aim of this letter is to introduce a covariant scheme for a
non minimal coupling between $GL(3,R)$ connection and the most
simple and non trivial class of matter fields in $2+1$ dimension,
without auxiliary fields involved, as a first step. The study of
$2+1$ dimensional gravity is a useful scenery to explore
interaction, taking in mind an extension to higher dimensions.

We demand the consistence condition that solutions, coming from
this type of theory must contains Einstein's gravity ones. At the
vacuum case, the last requirement means to consider the
quadratical power of cosmological constant. On the other hand,
when matter fields are turned on, it is observed that consistence
give rise constraints on matter fields and background. The last
step in this letter consists to include action terms dependent on
auxiliary fields, showing that some of restrictions can be
avoided.

Next we start with a brief review about notation. Let
${V_{\mu}}^a$ be the components of a tensorial object with curved
and lorentz indexes, defined in an $N$ dimensional space provided
with metric $g_{\mu \nu }$, curved coordinates
$x^\mu $, with $\mu ,\nu ,...=0,1,...,N-1$ and locally flat $\xi ^a $, 
with $a,b,c,...=0,1,...N-1$
(the Minkowski metric is $\eta _{ab} = diag(-1,+1,...,+1)$). Then,
introducing the spin (i.e., ${\omega _{\mu b}}^a $) and  affine
(i.e., ${\Gamma ^\lambda }_{\mu \nu }$) connections, the well
known covariant derivative is $ D_\mu {V_\nu }^a = \partial _\mu
{V_\nu }^a  +{\omega _{\mu b}}^a {V_\nu }^b - {\Gamma ^\lambda
}_{\mu \nu } {V_\lambda }^a $\,, etc. Particularly, one can demand
the property $D_\mu {e_\nu }^a =0$, where ${e_\nu }^a$ is the {\it
n-bein }which satisfies $g_{\mu \nu } = {e_\mu }^a {e_\nu }^b \eta
_{ab}$. From this, follows that the torsion can be written in the
form
\begin{equation}
{T^\lambda }_{\mu \nu } \equiv {\Gamma ^\lambda }_{\mu \nu } - {\Gamma
^\lambda }_{\nu \mu } = {e^\lambda }_a (\partial_\mu {e_\nu }^a - 
\partial_\nu {e_\mu }^a
+ {\omega _{\mu \nu }}^a - {\omega _{\nu \mu }}^a)
\, \, . \label{eq1}
\end{equation}

If the matrix elements for $GL(N,R)$ and lorentz transformations
are defined as ${(U)^\alpha }_\mu \equiv \frac{\partial {x^\prime
}^\alpha }{\partial x^\mu}$ and ${(L)^a}_b \equiv \frac{\partial
{\xi^\prime }^a }{\partial \xi^b}$, covariant behavior of
derivative $D_{\mu}$ demands the following transformation rules
for connections
\begin{equation}
{\omega ^\prime }_\mu  =L \omega _\mu L^{-1} + L \partial _\mu
L^{-1} \, \, , \label{eq2}
\end{equation}
\begin{equation}
{A_a}^\prime =UA_a U^{-1} + U \partial _a U^{-1} \, \, ,
\label{eq3}
\end{equation}
where we have introduced notation
\begin{equation}
{(A_a)^\mu }_\nu \equiv {e^\alpha }_a {\Gamma ^\mu }_{\alpha \nu }
\, \, . \label{eq4}
\end{equation}
It can be observed that the object (4) is a $GL(N,R)$ connection
which transforms like a lorentzian vector in flat index. So,
$GL(N,R)$ is chosen as the structure group, and will be assumed
that the fibre projection, the transition functions, etc. are
given. We do not discuss in this letter about non-compactness of
$GL(N,R)$ and its generators.

The Riemann-Christoffel curvature tensor (i.e.,
${R^\alpha}_{\sigma \mu \nu }$) is written through application of
the commutator $[D_\mu ,D_\nu]\equiv D_\mu D_\nu - D_\nu D_\mu $
on a rank one tensor, in other words $[D_\mu ,D_\nu]V^\alpha
={R^\alpha}_{\sigma \mu \nu } V^\sigma - {T^\lambda }_{\mu \nu }
D_\lambda V^\alpha $\,, and for all $V^\alpha $ says that
${R^\alpha}_{\mu \beta \delta } = \partial _\beta {\Gamma ^\alpha
}_{\delta \mu  } - \partial _\delta  {\Gamma ^\alpha }_{\beta \mu
} + {\Gamma ^\alpha }_{ \beta \lambda} {\Gamma ^\lambda }_{\delta
\mu }
-  {\Gamma ^\alpha }_{\delta \lambda} {\Gamma ^\lambda }_{\beta \mu  }$. 
Using this definition,
the Riemann-Christoffel tensor components can be written in the
form
\begin{equation}
{R^\alpha}_{\mu \beta \delta }=  {e_\delta}^a 
{e_\beta}^b{(F_{ab})^\alpha }_\mu
\, \,, \label{eq5}
\end{equation}
where
\begin{equation}
F_{ab} \equiv D_b A_a - D_a A_b + [A_a , A_b ]- {T^c }_{a b} A_c
\, \,, \label{eq6}
\end{equation}
is a Yang-Mills like curvature with torsion contribution.

In order to study the relationship between solutions obtained from
a Yang-Mills type lagrangian formulation defined
with curvature $F_{ab}$, and solutions of Einstein's gravity
with cosmological constant in a Riemannian space, we will consider
that Ricci tensor satisfies the field equation of Einstein (i.e.,
$ R^{\alpha \beta }-\frac{g^{\alpha \beta }}2R-\lambda g^{\alpha
\beta}=-8\pi G T^{\alpha \beta }$, where $T^{\alpha \beta }$ is
the energy-momentum tensor associated to material fields and
$\lambda$ is the cosmological constant).This ends the review.

We proceed noting the formulation without matter. Let the gauge
invariant action for $T^{\alpha \beta } =0$ be
\begin{equation}
S_o = \int d^N x \sqrt{-g} \,\, tr(-\frac 14   F_{\alpha
\beta}F^{\alpha \beta} + \Lambda (\lambda)) \,\, , \label{eq7}
\end{equation}
where $\Lambda (\lambda)$ is a constant that we expect it shall
depends on cosmological constant. One can see that action (7)
match with a pure Riemann-Christoffel quadratic lagrangian theory
in a torsionless space-time. Next, We will assume a Palatini's
variational principle type (i.e., $GL(N,R)$ connection and {\it
n-beins}
are our dynamical fields).
First, functional variation on connection $A$ gives
\begin{equation}
\delta _A S_o=\int d^N x \sqrt{-g}\,\,tr( E^\sigma \delta A_\sigma
) \,\, , \label{eq8}
\end{equation}
up to a boundary term. Notation means $
E^\sigma \equiv \frac 1{\sqrt{-g}}\,\,\partial _\mu 
(\sqrt{-g}\,\,F^{\sigma \mu }) + [F^{\sigma \mu } ,A_\mu ]
$\,, and in a torsionless space-time must be $ E^\sigma = D_\mu
F^{\sigma \mu } $. This last equation can be written in terms of
Ricci and Weyl (i.e., ${C_{\lambda \mu \nu }}^\alpha $) tensors
with the help of a Bianchi identity in the form
\begin{equation}
(E_\nu )_{\lambda \mu }= \frac {1}{(N-2)}(D_\lambda R_{\mu \nu }-D_\mu 
R_{\lambda \nu })
+ \frac {N-3}{N^2-3N+2} ( g_{\mu \nu }{\delta ^\rho }_\lambda
-g_{\lambda \nu }{\delta ^\rho }_\mu ) D_\sigma {R_\rho }^\sigma
+ D^\sigma {C_{\lambda \mu \sigma \nu }}
\,\,. \label{eq9}
\end{equation}

In $N=2+1$ dimension, the Weyl's tensor vanishes, and eq.(9) take
a simpler shape
\begin{equation}
(E_\nu )_{\lambda \mu } = D_\lambda R_{\mu \nu } - D_\mu R_{\lambda \nu }
\,\,. \label{eq10}
\end{equation}
An inspection on free field equation show that  de Sitter and Anti
de Sitter (i.e., dS/AdS) space are trivial solutions, among a huge
set of solutions. A similar fact shall occur in conformally flat
spaces in $N \geq 3+1$ dimensions.

Then, taking the trivial solution
$R^{\alpha \beta }=-2 \lambda
g^{\alpha \beta}$, the curvature $F_{ab}$ can be evaluated using
eq.(5):
\begin{equation}
{(F_{ab})^\alpha }_\mu = \lambda ({e^\alpha }_a e_{\mu b}-{e^\alpha }_b 
e_{\mu a})
\, \,, \label{eq11}
\end{equation}
which says the field $A_a$ is a pure gauge in dimension $2+1$ with
$\lambda =0$. This is the flat behavior that one expects in
vacuum. The topological sense of $\lambda =0$ would suggest
another interpretation for $F_{ab} =0$: is the field equation
coming from a Chern-Simons theory constructed with the connection
$A_a$.

Staying in a $2+1$ dimensional space, we consider variations of
the action $S_o$ on {\it dreibeins}. So, using eq.(11) we can
write
\begin{eqnarray}
\delta _e S_o \mid _{dS/AdS}=\int d^3 x \sqrt{-g}\,\,tr(\,F_{db}{F_c}^b 
- \frac 14 \eta _{dc}F^{ab}F_{ab}
+ \Lambda \,\eta _{dc}\,)\,\,{e^\sigma}^c \delta {e_\sigma}^d \mid 
_{dS/AdS} \nonumber \\
= \int d^3 x \sqrt{-g}\,(\Lambda -\lambda ^2 )\,{e^\sigma}_a
\delta {e_\sigma}^a \,\, , \label{eq12}
\end{eqnarray}
which says that the field equations for  dreibeins (joined with
the dS/AdS solutions of Yang-Mills field equations) fix the value
of unknown constant at $\Lambda = \lambda ^2$. In other words,
this condition guarantees that dS/AdS are trivial solutions for
the lagrangian formulation given in eq.(7).

Now, the model related to the coupling with matter through the
connection $A_a$ is outlined. We explore a possible covariant
scheme for non minimal coupling without auxiliary fields in the
following manner
\begin{equation}
S=S_o + \int d^3 x \sqrt{-g}\,tr(\ell(e,\psi) + 4\pi
G\,M^{ab}(e,\psi)F_{ab} )
\, \,, \label{eq13}
\end{equation}
where gauge invariant functional $\ell(e,\psi)$ and tensor
$M^{ab}(e,\psi)$ are functionals on dreibeins and material fields.
 From these definitions follows that, after a integration of action
(13) it can be obtained a ``current'' (i.e., minimal coupling) and
a ``mass'' (i.e., Proca type coupling) terms. Now, our problem is
to explore the shape of objects $\ell(e,\psi)$ and
$M^{ab}(e,\psi)$ requiring consistence with Einstein's solutions.

Here, we propose
\begin{equation}
{(M^{\alpha \beta})^\mu }_\nu = {(N^{\alpha \beta \sigma
\rho})^\mu}_\nu \,\,T_{\sigma \rho} + a{(n^{\alpha
\beta})^\mu}_\nu \, \,, \label{eq14}
\end{equation}
where $a$ is a real parameter, and objects $N^{\alpha \beta \sigma
\rho}$ and $n^{\alpha \beta}$ depend only on metric. Their general
form, in consistence with symmetry properties are
\begin{equation}
{(N^{\alpha \beta \sigma \rho})^\mu}_\nu \equiv g^{\mu \alpha }
{\delta ^\beta }_\nu g^{\sigma \rho} -g^{\mu \beta}{\delta ^\alpha
}_\nu g^{\sigma \rho}
+ g^{\rho \alpha } {\delta ^\beta }_\nu g^{\sigma \mu}-g^{\rho \beta}{\delta
^\alpha }_\nu g^{\sigma \mu}+ g^{ \mu \alpha } g^{\rho \beta
}{\delta ^\sigma }_\nu -g^{\mu \beta}g^{\rho \alpha }{\delta
^\sigma }_\nu
\, \,, \label{eq15}
\end{equation}
\begin{equation}
{(n^{\alpha \beta})^\mu}_\nu \equiv g^{\mu \alpha } {\delta ^\beta
}_\nu - g^{\mu \beta}{\delta ^\alpha }_\nu \, \,. \label{eq16}
\end{equation}

It is very important to note that in this first approach to a
coupling scheme we consider a system whose energy-momentum tensor
does not explicitly depend on connection. One can think in a class
of this type of systems. For example, the energy-momentum tensor
take the form
\begin{equation}
T_{\mu \nu } = (\alpha \,{\delta ^\lambda }_\mu {\delta ^\rho
}_\nu + \beta \,g^{\lambda \rho}g_{\mu \nu})\psi _{\lambda \rho}
\, \,, \label{eq17}
\end{equation}
with $\alpha $ and $\beta$ real scalars, and $\psi _{\lambda
\rho}$ be a symmetric tensor containing information about matter
fields. Tensor (17) can describe some interesting systems. If
$\alpha = 1$, $\beta = -\frac12$ and $\psi _{\lambda \rho}=
\partial _{\lambda}\phi \partial _{\rho}\phi$ we are talking about
a massless real scalar field $\phi$. On the other hand, if $\alpha
= p+\rho$, $\beta = p$, and $\psi _{\lambda \rho}= U_{\lambda}
U_{\rho}$, we are considering a perfect fluid with density $\rho$,
pressure $p$ and velocity $U_{\lambda}$. In a similar way, taking
adequate definitions can be included an electro-magnetic field or
a bosonic string. Any way, we assume that $\psi _{\lambda \rho}$
does not depend on metric nor connection (maybe some situation
where $\psi _{\lambda \rho}$ depends only on metric would be
considered, but the essentially physical  results will not be much
different).

So, the following step consists to performing $\delta _A S$
variations in eq.(13)
\begin{equation}
\delta _A S =\delta _A S_o - 8\pi G\int d^3 x \sqrt{-g}\,trD_\beta
M^{\sigma \beta}(e,\psi)\delta A_\sigma
\, \,, \label{eq18}
\end{equation}
up to a boundary term. When eq.(18) is evaluated on solutions of
Einstein gravity with cosmological constant  ($EG \lambda $), we
obtain
\begin{equation}
\delta _A S \mid _{EG \lambda } =4 \pi G\int d^3 x
\sqrt{-g}\big(g^{\sigma \lambda} D_\beta {T_\mu }^\beta -{\delta
^\sigma }_\mu D_\beta T^{\lambda \beta }\big)\delta
{(A_\sigma)^\mu }_\lambda
\, \,, \label{eq19}
\end{equation}
which says that $EG \lambda$ are extremals if the energy-momentum
tensor is a conserved one (sufficient condition).

But, as we will show there are more restrictions on material
fields when dreibeins variations are performed on action $S$.
Obviously, we need to say something about the shape of
$\ell(e,\psi)$. On one hand, this lagrangian density must be
consistent with vacuum limit of the theory ($\ell(e,\psi)
\rightarrow 0$ if $\alpha $ and $ \beta $ go to zero). On the
other hand, we demand consistency with {\it no gravity coupling
limit}, which consists to perform the limit $G \rightarrow 0$ and
$g_{\mu \nu}\rightarrow \eta_{\mu \nu}$ with $\lambda = 0$. Under
these conditions, action (13) must corresponds to a free flat
theory of matter fields. With this in mind, we propose a general
form for the matter fields action:
\begin{equation}
\int d^3 x \sqrt{-g}\,tr\,\ell(e,\psi)= S(\psi) +\int d^3 x
\sqrt{-g}\,\big(b_1 T^2 + b_2 T_{\mu \nu}T^{\mu \nu}\big)\equiv
S(\psi) +b \int d^3 x \sqrt{-g}\,{\psi}^2
\, \,, \label{eq20}
\end{equation}
where $S(\psi)$ is the action of matter fields in a curved
background and $b=b_1 (\alpha + 3\beta )^2 + b_2 (\alpha
^2+2\alpha \beta +3\beta ^2)$ is a real parameter. Soon we study
the no gravity coupling limit for $b$.

Then, variations of dreibeins in action (13) can be computed and
evaluated on $EG\lambda $, and a not too  large calculation gives
\begin{equation}
\delta _e S \mid _{EG \lambda } = \int d^3 x \sqrt{-g}\,{P^\sigma
}_d \big[\psi _{\alpha \beta}, {e^\mu}_b \big]\delta {e_\sigma}^d
\, \,, \label{eq21}
\end{equation}
where we introduce notation for a polynomial functional on $\psi
_{\alpha \beta}$ and dreibeins
\begin{eqnarray}
{P^\sigma }_d \big[\psi _{\alpha \beta}, {e^\mu}_b \big] \equiv
-(b+4\beta(8\pi G)^2(7\alpha +15\beta))\psi {\psi ^\sigma}_d \nonumber \\
+(b+(8\pi G)^2(-\alpha ^2 +2\alpha \beta + 9\beta
^2)){\psi}^2{e^\sigma}_d    \nonumber \\
+(-16\pi G(\alpha +4\beta)\lambda +(16\pi G)^2a\beta -\beta)\psi
{e^\sigma}_d \nonumber \\
+(48\pi G(3\alpha +10\beta)\lambda
+(16\pi
G)^2a\alpha -\alpha){{\psi}^\sigma}_d \nonumber \\
-32\pi Ga\lambda {e^\sigma}_d  \, \,. \label{eq22}
\end{eqnarray}

If one expect that action (13) must be an extremal on  $EG
\lambda$, equation of motion ${P^\sigma }_d =0 $ necessarily
represents a restriction for material fields. Therefore we assume
the consistent restriction
\begin{equation}
\psi =constant \,\,\epsilon \,\,R_e \, \,, \label{eq23}
\end{equation}
on possible matter field
configurations. This is not severe in the special case of a
perfect fluid (i.e., $\psi = U^\mu U_\mu =-1$). Anyway, for all
${\psi ^\sigma}_d$ with $\psi =constant $, equation $\delta _e S
\mid _{EG \lambda} =0$ gives two relations for $a$ and $b$
\begin{equation}
b \psi - a(16\pi G)^2 \alpha =-4\beta(8\pi G)^2(7\alpha
+15\beta)\psi +48\pi G(3\alpha  + 10\beta )\lambda -\alpha \, \,,
\label{eq24}
\end{equation}
\begin{equation}
b \psi ^2  + a 32\pi G(8\pi G \beta \psi -\lambda) =(8\pi
G)^2(\alpha ^2 -2\alpha \beta -9\beta ^2)\psi ^2 +(16\pi G(\alpha
+ 4\beta ) \lambda +\beta )\psi \, \,. \label{eq25}
\end{equation}

Solving this system for $b$, follows
\begin{eqnarray}
b=[(8\pi G)^3(\alpha ^3 -2{\alpha }^2 \beta-37\alpha \beta ^2
-60\beta ^3)\psi ^2+ (2(8\pi G)^2(\alpha ^2 +27\alpha \beta
+60\beta ^2)\psi
\nonumber \\
-48\pi G (3\alpha +10\beta )\lambda +\alpha )\lambda ][\psi (16\pi
G(\alpha +\beta )\psi - \lambda)]^{-1} \, \, ,\label{eq26}
\end{eqnarray}
with
\begin{equation}
\psi (16\pi G(\alpha +\beta )\psi - \lambda) \neq 0 \, \,.
\label{eq27}
\end{equation}
Parameter $a$ is obtained in a similar way.  Note that parameters
$b_1$ and $b_2$ remain undetermined.

It can be observed that, at no gravity coupling limit, eq.(26)
gives $b \approx (8\pi G)^2 (...)\longrightarrow 0$, as we have
expected (in other words, eq.(20) goes to the flat action of free
matter fields).

Equation (27) establishes the forbidden values for $\psi$ in
general. But, remembering the case of a perfect fluid in which
$\psi =-1$, this relation would represents a restriction for
cosmological constant (i.e., $\lambda \neq -16\pi G( \rho+2p)$).

We want to underline that eq.(23) and eq.(27)
means that the non minimal coupling scheme presented in eq.(13) is
consistent with Einstein gravity only under certain conditions
related with the class of matter distribution and the features of
space-time. This is not a surprising idea. In fact, from the point
of view of high spin gauge fields coupled to gravity it can be
found that theory is consistent only in restricted
backgrounds$^{3}$. So, maybe new interaction  terms (i.e.,
involving auxiliary fields ) added to action (13) enclose the hope
to reduce the number of constraints on matter fields.

In this sense, let $S'$ be the new action with $J^{a}$ as a some
functional on dreibeins and material fields, and
${(W_a)^{\mu}}_{\nu}$ the components of a auxiliary field that
transforms like a $GL(3,R)$ connection, then we write
\begin{eqnarray}
S'=S_o + \int d^3 x \sqrt{-g}\,tr\big(\ell(e,\psi) + 4\pi
G\,M^{ab}(e,\psi)F_{ab}  \nonumber \\
+J^{a}(A_a-W_a)+H^{a b}(A_a-W_a)(A_b-W_b)\big)
\,\, \,, \label{eq28}
\end{eqnarray}
where a naive proposal for components of $J^{a}$ and $H^{a b}$ is
taken
\begin{equation}
(J_{\beta})_{\mu \nu }\equiv (d_1 + d_2 \psi){\varepsilon }_{\beta
\mu \nu} \, \,, \label{eq29}
\end{equation}
\begin{equation}
( H^{\alpha \beta})^{\mu \nu} \equiv a_1
\,g^{\alpha\beta}g^{\mu\nu}+a_2 \,g^{\alpha\mu}g^{\beta\nu}+ a_3
\,g^{\alpha\nu}g^{\beta\mu}
\, \,, \label{eq30}
\end{equation}
with the real parameters $d_1$, $d_2$, $a_1$, $a_2$ and $a_3$.

Variations on $W_b$ give $\frac{\delta _{W}S'}{\delta W_b}=-\int
d^3 x \sqrt{-g}\,tr\big(J^{b}+H^{a b}(A_a-W_a)+(A_a-W_a)
H^{ba}\big)$. For all $\delta W_a$, equation of motion is
\begin{equation}
J^{b} + H^{a b}(A_a-W_a)+(A_a-W_a) H^{ba}=0
\, \,. \label{eq31}
\end{equation}
This says that dynamic of connection fields is maintained (i.e., $
\delta _A S' \mid _{EG \lambda \, , \,\delta _{W}S'=0}=\delta _A S \mid 
_{EG \lambda }=0
$). On the other hand, suggests an ansatz for auxiliary fields:
\begin{equation}
(A_{\alpha}-W_{\alpha})_{\mu \nu}= (\theta _1 +  \theta _2 \psi )
\,{\varepsilon }_{\alpha\mu\nu}\, \,, \label{eq32}
\end{equation}
with $\theta _1$ and $\theta _2$ real parameters (this is not the
most general linear dependence on field $\psi _{\sigma \beta}$,
but it is sufficient and consistent with eq.(31)).

Next, arbitrary variations on dreibeins for $\delta _{e}S'\mid
_{EG \lambda \, , \,\delta _{W}S'=0}=0$ produce
\begin{eqnarray}
{P^\sigma }_d \big[\psi _{\alpha \beta}, {e^\mu}_b \big] + tr
\frac{\delta J^{\beta}}{\delta
{e_\sigma}^d}(A_{\beta}-W_{\beta})+tr \frac{\delta H^{\alpha
\beta}}{\delta
{e_\sigma}^d}(A_{\alpha}-W_{\alpha})(A_{\beta}-W_{\beta})\nonumber \\
-tr H^{\alpha \beta}(A_{\alpha}-W_{\alpha})(A_{\beta}-W_{\beta})
\,\,{e^\sigma}_d=0\, \,, \label{eq33}
\end{eqnarray}

Then, using eq.(22) and eq.(32) in eq.(33), it can be found for
any material field an equation system for free parameters
\begin{equation}
16\pi G a \lambda + 9(a_1 -  a_2){\theta _1}^2=0 \,, \label{eq34}
\end{equation}
\begin{equation}
16\pi G(\alpha +4\beta)\lambda -(16\pi G)^2a\beta +\beta +36(a_1 -
a_2)\theta _1\theta _2=0 \,, \label{eq35}
\end{equation}
\begin{equation}
48\pi G(3\alpha +10\beta)\lambda +(16\pi G)^2a\alpha -\alpha
+48(a_1 - a_2)\theta _1\theta _2=0 \,, \label{eq36}
\end{equation}
\begin{equation}
b+(8\pi G)^2(-\alpha ^2 +2\alpha \beta + 9\beta ^2)-18(a_1 -
a_2){\theta _2}^2=0 \,, \label{eq37}
\end{equation}
\begin{equation}
b+(8\pi G)^2(28\alpha \beta + 60\beta ^2)+48(a_1 - a_2){\theta
_2}^2=0 \,, \label{eq38}
\end{equation}
which is only consistent for $a$, $b$, $(a_1 -  a_2){\theta
_1}^2$, $(a_1 -  a_2){\theta _2}^2$ and $\lambda$ as unknown
parameters for a given $\alpha$ and $\beta$. From these equations,
a restriction for cosmological constant can be obtained. For
example, in the massless scalar field case, this says that
$\lambda =\frac{5}{192\pi G}> 0$ (i.e., De Sitter background).

We finalize, saying that although the coupling presented in
eq.(28) has not the most general form, the idea in that strong
restrictions on ${\psi }_{\alpha \beta}$ can be avoided with
auxiliary fields has been elucidated.

Author wish to thanks Prof. P.J. Arias
for interesting discussions and
observations. Also thanks Prof. F. Mansouri for references.

\end{document}